\documentclass[reprint,amsmath,amssymb,aps,pra,superscriptaddress]{revtex4-2}
\usepackage{mathrsfs}
\usepackage{amsthm}
\usepackage{graphicx}
\usepackage{bm, bbm, physics}
\usepackage{graphics,graphfig,graphicx,physics,setspace,indentfirst,graphicx,lipsum} 
\usepackage{amsmath,amsthm,amsfonts,amssymb,amscd,enumerate,mathrsfs, bbm}
\usepackage{subfig, float}
\usepackage{tikz}
\usepackage{hyperref}
\usepackage{caption, ragged2e}
\usepackage{qcircuit}
\usepackage{array}
\newcolumntype{C}[1]{>{\centering\arraybackslash}m{#1}}
\newcommand{\T}{\mathcal{T}}
\newcommand{\SSS}{\mathcal{S}}
\newcommand{\SM}{\mathcal{S}_{\text{Mat}}}
\newcommand{\OO}{\widetilde{O}(\theta)}

\begin{document}

\title{Symmetry-guided gradient descent for quantum neural networks}

\author{Kaiming Bian}
\affiliation{Shenzhen Institute for Quantum Science and Engineering, Southern University of Science and Technology, Nanshan District, Shenzhen, China}
\affiliation{Department of Physics, Southern University of Science and Technology, Nanshan District, Shenzhen, China}

\author{Shitao Zhang}
\affiliation{Shenzhen Institute for Quantum Science and Engineering, Southern University of Science and Technology, Nanshan District, Shenzhen, China}

\author{Fei Meng} 
\email{feimeng@cityu.edu.hk}
\affiliation{Department of Physics, City University of Hong Kong, Tat Chee Avenue, Kowloon, Hong Kong SAR}

\author{Wen Zhang}
\email{zhangwen20@huawei.com}
\affiliation{HiSilicon Research, Huawei Technologies Co., Ltd., Shenzhen, China}

\author{Oscar Dahlsten}
\email{oscar.dahlsten@cityu.edu.hk}
\affiliation{Department of Physics, City University of Hong Kong, Tat Chee Avenue, Kowloon, Hong Kong SAR}
\affiliation{Shenzhen Institute for Quantum Science and Engineering, Southern University of Science and Technology, Nanshan District, Shenzhen, China}
\affiliation{Institute of Nanoscience and Applications, Southern University of Science and Technology, Shenzhen 518055, China}

\date{\today}

\begin{abstract}
Many supervised learning tasks have intrinsic symmetries, such as translational and rotational symmetry in image classifications. These symmetries can be exploited to enhance performance. We formulate the symmetry constraints into a concise mathematical form. We design two ways to adopt the constraints into the cost function, thereby shaping the cost landscape in favour of parameter choices which respect the given symmetry. Unlike methods that alter the neural network circuit ansatz to impose symmetry, our method only changes the classical post-processing of gradient descent, which is simpler to implement. We call the method symmetry-guided gradient descent (SGGD). We illustrate SGGD in entanglement classification of Werner states and in {two classification tasks} in a 2D feature space. In both cases, the results show that SGGD can accelerate the training, improve the generalization ability, and remove vanishing gradients, especially when the training data is biased. 
\end{abstract}

\maketitle

\textit{Introduction.}---Quantum machine learning extends concepts from classical machine learning into the regime of quantum superposition~\cite{schuld2015introduction, cerezo2022challenges}. 
Quantum neural networks can be viewed as quantum generalizations of classical neural networks, amounting to an extension of deep learning to the quantum regime~\cite{li2022quantum, wan2017quantum, romero2017quantum}. QNNs have shown promising results, for example, in quantum phase recognition~\cite{cong2019quantum} and classical classification tasks~\cite{perez2020data}.
\begin{figure}[t!]
    \centering
    \includegraphics[width = 8.6 cm]{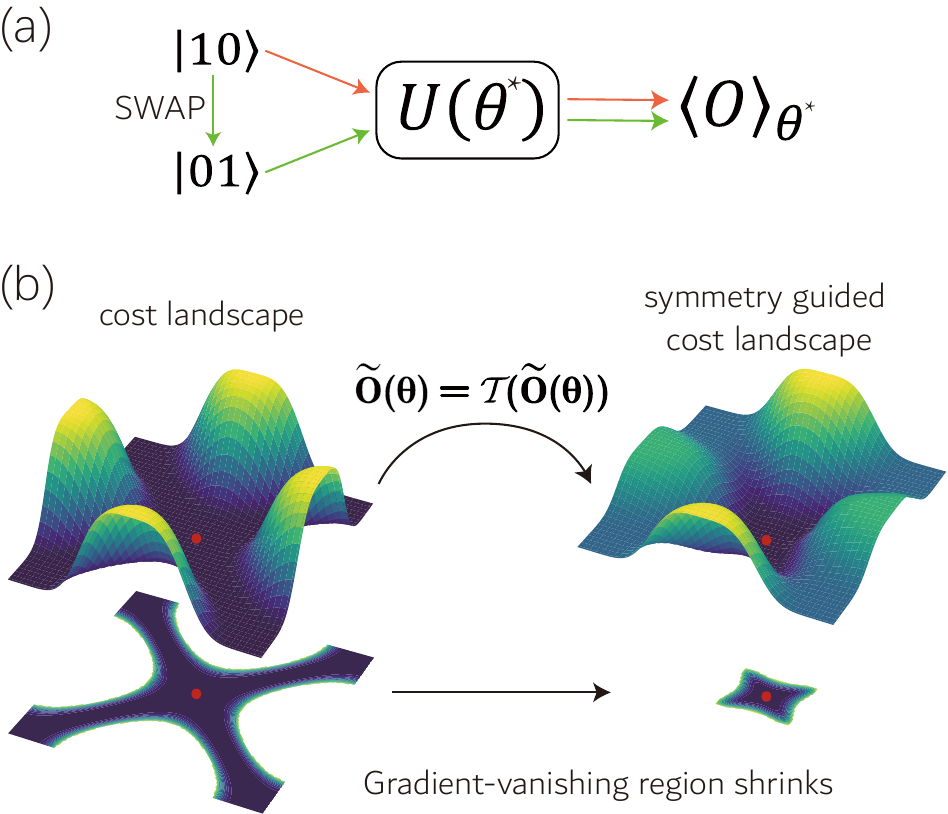}
    \caption{
    \justifying
    {\bf Imposing symmetry of a task on the cost landscape.} (a) Suppose $\ket{1,0}=\mathrm{SWAP}\ket{0,1}$ and $\ket{0,1}$ should have the same label so that SWAP is a task symmetry. A circuit with optimal parameter $\theta^*$ gives output $\expval{O}_{\theta^*}$ determining the assigned label. 
    (b) The neural net circuit is trained via gradient descent on a cost landscape. Using the symmetry constraint equation~\eqref{eqn: symmetry constraint equation}, we shape the cost landscape given the symmetry of the data labels. In this example, the gradient-vanishing region around the optimal parameter $\theta^*$ then shrinks, where $\theta^*$ is represented by the red dot. The symmetry guidance method moreover improves training performance given biased data.
    }
    \label{fig:illustrative}
\end{figure}
QNNs employ parameterized quantum circuits (PQC) ~\cite{wan2017quantum, benedetti2019parameterized, vidal2018calculus,sim2019expressibility}, with a classical optimizer updating the parameters during training. There are now several variants including quantum graph neural networks~\cite{verdon2019quantum, ai2022decompositional}, quantum convolution neural networks~\cite{cong2019quantum,oh2020tutorial}, and tensor network-based QNNs~\cite{huggins2019towards, chen2020hybrid, sun2020generative}.

Leveraging symmetry can improve the performance of both classical and quantum neural networks.  
Inspired by successful machine learning models, a field known as geometric deep learning investigates the relation between symmetry and learning~\cite{bronstein2017geometric, bronstein2021geometric}. Recently, such ideas have been extended into the quantum regime, establishing the field of geometric quantum machine learning~\cite{larocca2022group,sbahi2022provably, mernyei2022equivariant, meyer2023exploiting}. To leverage symmetry one may modify the quantum circuits being trained to gain significant performance improvements~\cite{meyer2023exploiting, lyu2023symmetry, seki2020symmetry}. However, the symmetry-preserving modification entails altering the PQC ansatz, which can be experimentally costly. In specific scenarios (see 
Appendix~\ref{appendix:Hardware inefficiency of circuit ansatz symmetrization})
such modifications of quantum circuits even transform local unitaries into global unitaries, {which are difficult to implement in experiments}. 
Such potential hardware inefficiencies motivated us to explore further methods to leverage symmetry properties to optimize the performance of quantum neural nets.

We give a mathematically justified method to modify the cost function to guide the parameters toward the space with the desired symmetry, a method which we call \textit{symmetry-guided gradient descent} (SGGD). {We formulate the symmetry constraints as a neat equation so that the constraints can easily be rewritten as a penalty term.} Specifically, we rewrite the symmetry constraint into a concise equation and modify the cost function to effectively guide the optimization of parameters, steering the quantum circuit to satisfy the symmetry. We designed two ways to implement SGGD, which involve averaging an observable with a symmetry group, a process called `twirling'. In one approach, the cost function gains a penalty term that suppresses the appearance of symmetry-breaking circuits by twirling. In the other approach, the original observable is replaced with the twirled one when calculating the cost function.  

We illustrate SGGD in numerical experiments with both quantum and classical inputs. For the case of quantum input, we consider the entanglement classification of Werner states~\cite{werner1989quantum}. SGGD is shown to prevent biased sampling of input states from leading to poor generalization, dramatically reducing the required training data size. {Additionally, it sharpens the slope of the cost curve which leads to faster training}. For the case of classical inputs, we consider classifying a set of 2D points with rotational symmetry into a few categories. The 2D data are encoded into quantum states by rotation gates~\cite{schuld2021machine, weigold2020data}. The encoded states with their labels are fed into a PQC for training. SGGD increases the accuracy from $69.9\%$ to $89.6\%$ {in classification into $2$ classes, and increases the accuracy from $49.5\% $ to $66.1\%$ in classification into $3$ classes} when the training data is highly biased. {In both cases, our method can improve the generalization ability significantly, especially when confronted with biased training data---a prevalent challenge within high-dimensional spaces~\cite{bronstein2021geometric, biwei2021dataaug}.
}

Our method, SGGD, is widely applicable to gradient descent over QNNs and other PQCs. The SGGD approach of changing the cost function is more convenient than altering the PQC ansatz because the modification is implemented classically, which does not impose additional workload on quantum devices. As demonstrated in the numerical examples, SGGD successfully reduces the occurrence of circuits that do not satisfy the symmetry, thereby improving the efficiency of the QNN model. 


\textit{QNN set-up.--} 
In QNN training, the PQC processes input data on a quantum chip while the classical optimizer runs on a classical computer to update parameters~\cite{cerezo2021variational, endo2021hybrid}.
The input data needs to be encoded into the quantum state if it is classical~\cite{schuld2018supervised, schuld2021machine, schuld2021supervised}. After the state is evolved by the PQC $U(\theta)$, commonly the output of QNN is determined by measuring an observable $O$ and obtaining its expectation value
\begin{equation}
    f_\theta(x) = \tr(U(\theta) \rho(x) U^\dagger(\theta) O),
\label{eqn: PRL_QNNForm}
\end{equation}
where the encoding $\rho$ maps classical data $x$ into the density matrix $\rho(x)$. Now, a classification function $h$, as classical post-processing, is applied to map this expectation value to the predicted label. For instance, a step function $h$ may be employed by the model to predict the label of $x$; if $f_\theta(x) \geq 0$, the model predicts the label of $x$ as $1$; otherwise, it predicts the label of $x$ as $-1$.

\textit{Symmetry constraint equation.--} The data label in many supervised learning tasks is invariant under certain symmetries.
For example, the number of $1$s in a bit string remains unchanged through swapping two bits (see Fig.~\ref{fig:illustrative} (a)). 
Ref.~\cite{bronstein2021geometric} gives a formal expression of an $\SSS$-invariant data label:
\begin{equation}
     f(s(x)) = f(x),~\forall x \in \mathcal{X},~ \forall s\in \SSS,
     \label{eqn: PRL 2}
\end{equation}
where $\SSS$ denotes the group that consists of symmetry operations. We describe the case where the classical post-processing function $h$ is identity first before examining the case of general $h$. 

If an objective function $f_\theta$ could approximate the target function $f$, there is an optimal parameter $\theta^*$ such that $f_{\theta^*}(s(x)) = f_{\theta^*}(x) + \epsilon$, where $\epsilon$ is a tolerable error. Combining this condition with Eq.~\eqref{eqn: PRL_QNNForm} and Eq.~\eqref{eqn: PRL 2}, the optimal PQC $U(\theta^*)$ that satisfies the symmetry of the data should obey
\begin{equation}
    \tr(\rho(x) S^\dagger\widetilde{O}(\theta^*)S ) = \tr(\rho(x) \widetilde{O}(\theta^*)) + \epsilon,
    \label{eqn: PRL 3}
\end{equation}
for all $S$ in a symmetry matrix group, where $\OO$ is $ U^\dagger(\theta) O U(\theta)$. It can be observed that, as long as the encoded states $\{\rho(x)\}$ do not strictly reside within a linear subspace of the entire Hilbert space $\mathcal{H}$, Eq.~\eqref{eqn: PRL 3} is equivalent to the following symmetry constraint equation (for the detailed derivation, see Appendix~\ref{appendix:Symmetry constraint equation})
\begin{equation}
    \T(\widetilde{O}(\theta^*)) = \widetilde{O}(\theta^*)+E,
    \label{eqn: symmetry constraint equation}
\end{equation}
where $E$ is a tolerable-size error matrix which satisfies $\norm{E}<\epsilon$. $\T$ is the twirling operator, which maps an operator $O$ into the symmetry-twirled Hamiltonian
\begin{equation}\label{eq:twirling}
    \T(O) =\int d\mu(S) S O S^\dagger,
\end{equation}
where $\mu$ is the Haar measure.

\textit{Symmetry guided gradient descent (SGGD).--}
Gradient descent uses the gradient of the cost function as the direction for updating parameters. A general method for guiding gradient descent is to include an extra penalty term~\cite{rossi2006handbook}. We say a gradient descent is guided by symmetry if the cost function is modified to make the parameters satisfy the symmetry constraint equation Eq.~\ref{eqn: symmetry constraint equation}. We now describe how to create a penalty term for a given symmetry.

We take the mean square error (MSE) as a default cost function $c_0(\theta)$, and construct a penalty term $g(\theta)$ that employs the Hilbert-Schmidt norm to evaluate the difference between $\OO$ and $\T(\OO)$,
\begin{align}
    & c_1(\theta) = c_0(\theta) + \lambda g(\theta) \notag \\
                =& \frac{1}{|\mathcal{X}|} \sum_{x\in \mathcal{X}} (f_\theta(x) - f(x))^2 + \lambda \norm{\T(\OO) - \OO},
\label{eq:cost1_add}
\end{align}
where $\mathcal{X}$ is the data set and $\lambda$ controls the intensity of the penalty term. 
The penalty term $g(\theta)$ would reach the minimum value $0$ when the circuit $U(\theta)$ respects the symmetry constraint equation. Hence, the $g(\theta)$ guides the parameters toward the symmetry-preserving space during gradient descent. As illustrated in Fig.~\ref{fig:illustrative}(b), the gradient vanishing region shrinks after including the penalty term in the cost function, implying an improvement in parameter training. 
Moreover, $g(\theta)$ aids in discovering the optimal $\theta$ irrespective of the quality or the quantity of data as the penalty constrains $\theta$ to the region with symmetry, compensating for the lack of suitable training data. 

{
The SGGD method could be directly applied to other gradient based methods, such as  ADAM~\cite{kingma2014adam} and mirror gradient descent~\cite{sbahi2022provably}, by adding the penalty term to their cost function. The performance can be further improved by tuning the intensity coefficient $\lambda$. Thus, SGGD can be added to gradient based methods rather than being mutually exclusive.
}

As a second way to incorporate symmetry guidance, we directly substitute the symmetry constraint equation into the objective function $f_\theta$, 
\begin{equation}
    c_2(\theta) = \frac{1}{|\mathcal{X}|} \sum_{x\in \mathcal{X}} \left( \tr(\rho(x) \T(\OO)) - f(x) \right)^2.
\label{eq:cost2_twirled}
\end{equation}
This cost function is guided by symmetry in the sense that only those parameters that lead to variations in $\T(\OO)$ can change the overall cost value $c_2$. 

The two methods differ in terms of what exactly is optimised and in their computational cost. $c_1$ only cares about preserving symmetry rather than predicting accuracy, while $c_2$ cares both about symmetry and the accuracy of the QNN. However, $c_2$ introduces more burden in terms of quantum devices since it makes explicit use of the twirled observable $\tilde{O}$~\cite{toth2007efficient} whereas $c_1$ can be evaluated via a swap test between the twirled and non-twirled observables. Whilst the twirled quantities can be calculated efficiently with a quantum device, in the numerical experiments in this paper we calculate $c_1$ and $c_2$ via naive and inefficient classical simulation. Whether calculated with quantum devices or classical methods, the penalty term only needs to be calculated once per training epoch without looping over training inputs.  

To illustrate and explain how and when SGGD is useful, we analyze two examples. We choose examples that are non-trivial yet sufficiently simple to also be analyzed by hand so one can gain a full understanding of when and why SGGD works. In some cases, the symmetry is an invariance of the classical label and in some cases the symmetry is an invariance of the quantum output under a unitary operation. The first example we consider is of the latter type and the second example is of the former. 

\textit{Example: entanglement classification.--}We now analyse the task of entanglement classification of 2-qubit Werner states. This example illustrates how our methods alleviate bad performance caused by biased training data. 

Consider the task to classify Werner states~\cite{dur2000classification, gao2017efficient} as entangled or separable. Two-qubit Werner states can be expressed as 
\begin{equation}
    \rho_{\mathrm{W}}=\frac{2-p}{6}I+\frac{2p-1}{6}\text{SWAP},
\end{equation}
where the parameter $p$ ranges from $-1$ to $1$. 
SWAP is a unitary and Hermitian matrix defined by $\text{SWAP}\ket{i,j}=\ket{j,i}$. Werner states are entangled for $p<0$ and separable for $p\ge0$~\cite{unanyan2007decomposition}. 
We take the demanded symmetry matrix group to be all tensor products of single-qubit unitaries, $U\otimes U$, because Werner states are invariant under these operations.

\begin{figure}[h]
    \centering
    \includegraphics[width = 8.6 cm]{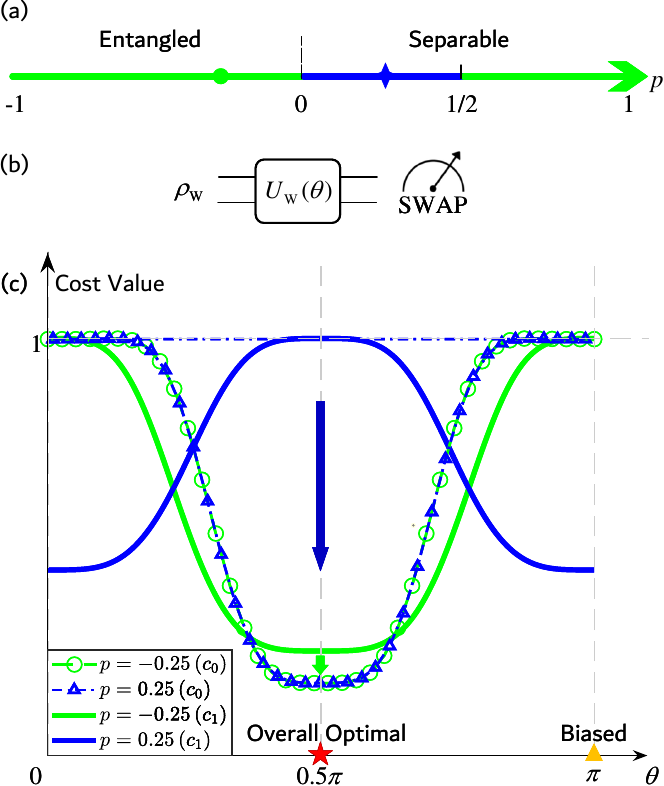}
    \caption{
    \justifying
    Settings and loss landscapes comparison for the classification of 2-qubit Werner states. (a) illustrates how the entanglement of Werner states changes along with the parameter p; states in different color areas have different kinds of loss landscapes in our case. (b) shows the QNN structure being used. (c) shows the loss curves before ($c_0$) and after ($c_1$) applying symmetry guidance for two representative samples from each color area, with $p=-0.25\text{ and }0.25$, respectively. Curves in this figure are rescaled, with each value divided by the maximum value of its corresponding curve.}
    \label{fig:prl_werner}
\end{figure}

Consider how to employ a QNN for the above task. Without loss of generality, we say a state is classified as entangled if the output of the QNN satisfies $f_\theta(\rho_{\mathrm{W}})< 0$ and as separable otherwise. The simplest QNN that is composed of the identity circuit followed by measuring the SWAP operator as the observable will output  $f_\theta(\rho_{\mathrm{W}})=p$, and thus classify the state correctly. For simplicity and clarity, we fix the observable as SWAP and parametrize the circuit as a unitary $U_W(\theta)$, where $\theta$ is the variational parameter to be trained. We assume a one-dimensional parameter as it helps to visualize the loss landscape easily. We use the following QNN ansatz, 
\begin{equation}
    U_\mathrm{W}(\theta) = {\rm{CNOT}}^{{{\cos }^2}(\theta)},
\end{equation}
which is simple enough to elaborate the mechanisms of SGGD while containing the unique optimal solution (identity) and a biased solution (CNOT). The ansatz structure is depicted in Fig.~\ref{fig:prl_werner}(b).
The cost landscape highly depends on the training data chosen. As shown in Fig.~\ref{fig:prl_werner}, we find that the whole data set can be split into two subsets $A=\{p\mid p\in [0,1/2) \}$ and $\Bar{A}=[-1,1]- A$. Training data from $\Bar{A}$ leads to a convex cost landscape that gives the correct solution under gradient descent, and training data from $A$ yields a concave landscape that gives the biased solution. One can verify the above claim by depicting the cost landscape for each data point in $A$ and $\Bar{A}$ respectively, as shown in Fig.~\ref{fig:prl_werner}(c) with $p=-0.25$ and $p=0.25$ as examples.

We find that SGGD alleviates the bias caused by training data in $A$. The symmetry group $G$ of input Werner states is $\{ U\otimes U \mid U \text{ is unitary} \}$. The penalty term given by Eq.~\eqref{eq:twirling} is
\begin{align}
    \mathcal{T}(\OO) = \alpha_1 I + \alpha_2 \text{SWAP},
    \label{eq:TwirledOperatorWerner}
\end{align}
where $\OO = U^\dagger_\mathrm{W}(\theta) \text{SWAP} U_\mathrm{W}(\theta)$, $\alpha_1 = \cos^2(\frac{\eta}{2})/2$, $\alpha_2 = \sin^2(\frac{\eta}{2})$, and $\eta = \pi \sin^2\theta$. Substituting the twirled operator of Eq.~\ref{eq:TwirledOperatorWerner} into Eq.~\eqref{eq:cost1_add} we get the penalty term for $c_1$ as
\begin{align}
    g(\theta) = \frac{3}{4}\Big( \sin^2 \eta + 2\cos \eta +2  \Big).
\end{align}
Note that minimizing the penalty alone could optimize the parameter $\theta$ since $g(\pi/2) = 0$ is the minimal value, and $\theta=\pi/2$ gives the optimal QNN circuit.

Using the SGGD cost function $c_1$,  all data points, no matter in $A$ or $\Bar{A}$, can give the optimal solution, as shown by the two dotted curves in Fig.~\ref{fig:prl_werner}(c). This implies that even if the majority of the training data is picked from $A$, QNN can still be trained correctly. SGGD thus dramatically improves the generalization performance of the QNN trained by biased data. Additionally, the cost curves of data from $\Bar{A}$ become steeper, and the training converges faster, indicating that SGGD can accelerate the training of QNNs. Experimental details and alternative circuit ansatzes are given in the Appendix~\ref{appendix:Entanglement classification}.

\begin{figure}[t]
    \centering
    \includegraphics[width = 8.6 cm]{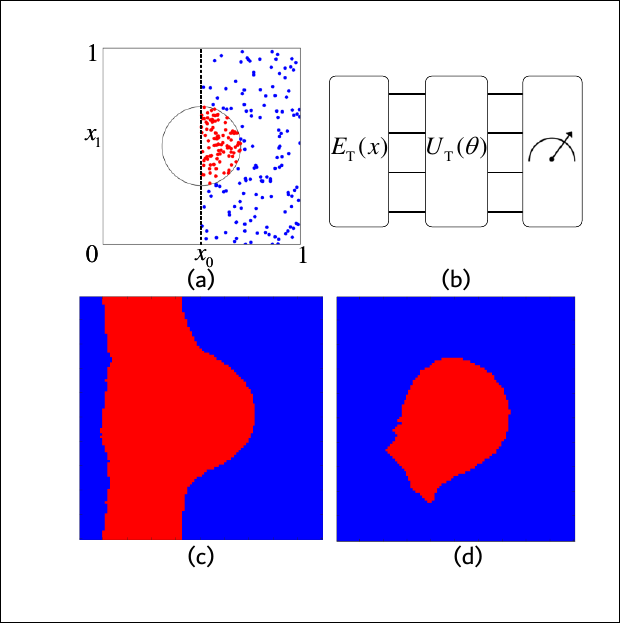}
    \caption{
    \justifying
    { {\bf  SGGD enhances 2D classical data classification with 2 classes using rotation encoding.} Panel (a) illustrates the distribution of the training data, which is highly biased. The data points are classified into two categories marked as red and blue. Panel (b) illustrates the QNN structure, where the rotation encoding  $E_T(x)$ and circuit ansatz $U_T(\theta)$ are detailed in the Appendix~\ref{appendix:circuit ansatz}. Panels  (c) and (d) demonstrate the predictions on test data without using SGGD (cost function $c_0$) and using SGGD (cost function $c_2$)}, respectively.
}\label{fig:prl_twoDim}
    
\end{figure}

\textit{Example: classification of 2-dimensional classical data.--} Consider the task of classifying classical data points on a 2D plane~\cite{hubregtsen2021evaluation} into two classes using hardware-efficient PQCs~\cite{kandala2017hardware}. As shown in Fig.~\ref{fig:prl_twoDim}(a), training data are sampled in the range $[0,1]^{2}$, and data points are classified into two categories depending on if their distance to the center point $(\frac{1}{2},\frac{1}{2})$ is smaller than $0.2$ or not (a case of non-linear classification). 
This task can be solved by QNNs with high accuracy when the training data are sampled uniformly from the whole data space~\cite{hubregtsen2021evaluation}. However, models can exhibit poor generalization performance when training samples are not uniformly sampled, e.g., when data is sampled from only the right half of the data space, as shown in Fig.~\ref{fig:prl_twoDim}(a). We choose one of the effective QNNs in Ref.~\cite{hubregtsen2021evaluation}, whose circuit structure is detailed in the Appendix~\ref{appendix:circuit ansatz}, and compare its prediction performances using the cost function $c_0$ and the SGGD cost function $c_2$.  SGGD with cost function $_2$ increases the accuracy from $69.9\%$ to $89.6\%$. The reason we choose $c_2$ here is that $c_1$ is overly restrictive, imposing symmetry at the level of the quantum output before the classical post-processing. We do not want the training to rule out models that respect the symmetry of the label but are not symmetric before the post-processing.

\begin{figure}[t]
    \centering
    \includegraphics[width = 8.6 cm]{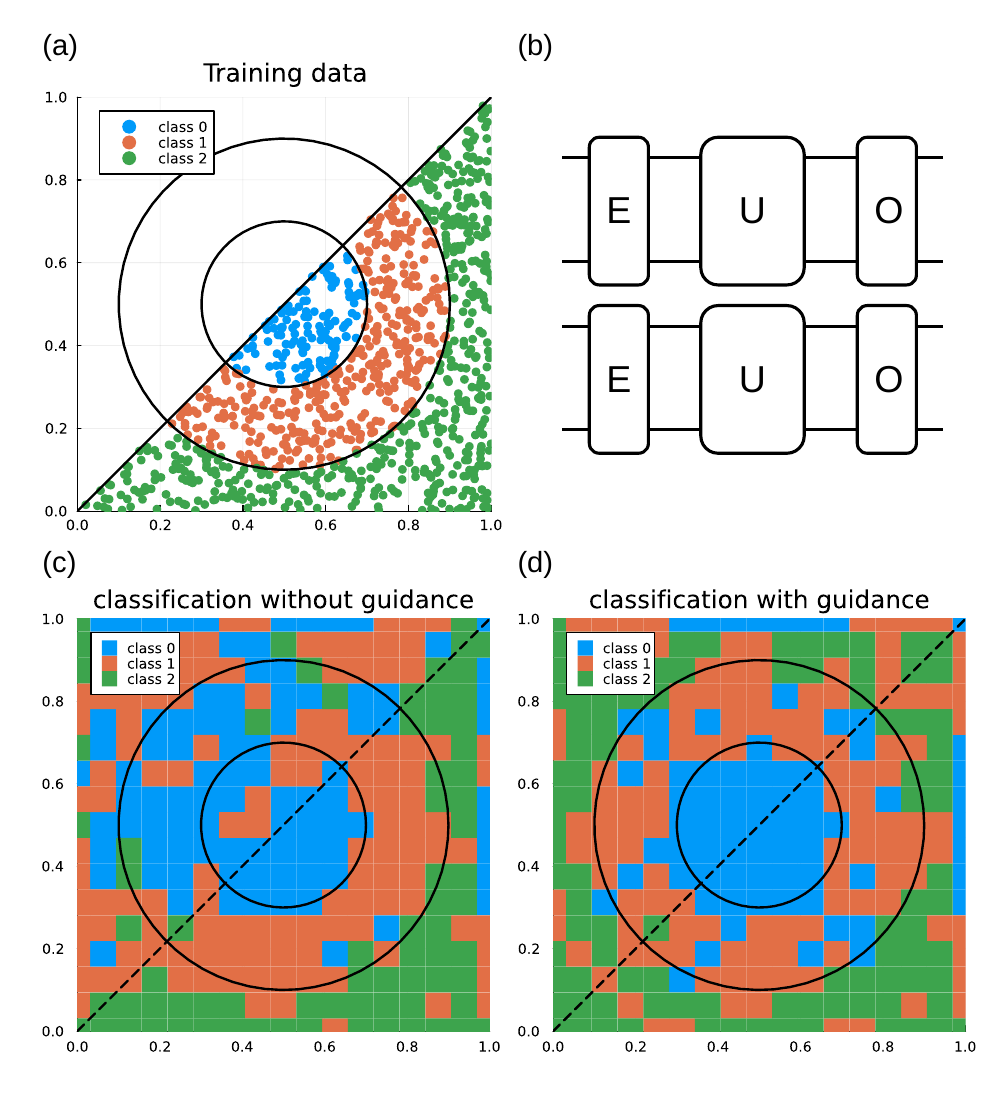}
    \caption{
    \justifying
    {
        {\bf  SGGD enhances 2D classical data classification with 3 classes using binary encoding}. Panel (a) illustrates the highly biased distribution of the training data set. Points are labeled into $3$ classes according to their relative position to the circles. (b)  Two identical circuit ansatzes for 2-class classification are applied to classify whether (i) the input data is inside the inner circle (the upper branch)  and (ii) inside the outer circle (the bottom branch). Each circuit branch has 8 qubits.  We use a binary encoder $E$ and the hardware-efficient ansatz $U$. Panel  (c) and (d) demonstrate the predictions on test data without using SGGD (cost function $c_0$) and using SGGD (cost function $c_2$), respectively.}
    }\label{fig:prl_reup}
    
\end{figure}

{
{SGGD also performs well on a generalization to the above 2D classification task with three classes and a larger quantum circuit with 16 qubits, as shown in Fig.~\ref{fig:prl_reup}. The data points in $[0,1]^2$ with coordinates $(x,y)$ are to be classified into three categories based on their distances $r$ to the center point $(0.5,0.5)$. If $r<0.2$, then the data point belongs to class 0; if  $r>0.4$, then it belongs to class 2; otherwise, it belongs to class 1. This classification can be solved by applying the 2-class classifier we obtained above twice; the first classifier tells whether the point is located inside the inner circle, and the second classifier outputs whether the point is located inside the outer circle. Via binary encoding, each coordinate $x$ or $y$ is truncated into a 4-digit binary string $\textbf{x}$ or $\textbf{y}$ and is represented by $4$ qubits respectively, resulting in a 8-qubit state $\ket{\textbf{x},\textbf{y}}$ that encodes a truncated data point. Then, we generate two copies of the state $\ket{\textbf{x},\textbf{y}}^{\otimes 2}$ using 16 qubits, where each copy is fed into the 2-class classifier quantum circuit, as shown in Fig.~\ref{fig:prl_reup}(b). The upper quantum circuit outputs whether the data point is inside the inner circle or not, and the bottom circuit outputs whether the data point is inside the outer circle or not. Combining the classification outcome of these two parallel circuits, we can classify the data points into one of the three classes. 

With highly biased training data shown in Fig.~\ref{fig:prl_reup}(a), the accuracy of prediction is only 49.5\%, while using SGGD to modify the cost function increases the accuracy to 66.1\%.
}
}

\textit{Non-trivial classical post-processing.--}We now consider non-trivial classical post-processing where $h$, a classical function of the output $f$, is not the identity. 
 
In this case, the requirement for a symmetry-invariant circuit changed from Eq.~\eqref{eqn: PRL 3} to $h(f_\theta( s(x) )) = h(f_\theta( x ))$.
Circuits that adhere to this equation are not necessarily bounded by the symmetry constraint equation (see details in the Appendix~\ref{appendix:Symmetry constraint equation}). For instance, consider the labels assigned to $\ket{01}$ and $\ket{10}$, both of which are assigned a label of $1$. However, their respective probabilities of yielding the label $1$ may differ.
If we force the parameters or the circuit to obey the symmetry constraint equation, their probabilities to get label $1$ are forced to be the same, which is overly restrictive and thus harms the expressive power of the PQCs. 

We propose a modified version of the symmetry constraint equation that integrates classical post-processing. The basic idea of modifying the symmetry constraint equation is considering $h$ as an analytical function that can be expanded using a Taylor series. By truncating the series to the $t$-order polynomial $h_t$, the effect of the post-processing could be formulated into a neat form,
\begin{equation}
    h_t (f(x)) = \tr\left( \Big(U(\theta) \rho(x) U^\dagger(\theta)\Big)^{\otimes t} O_{h_t}(\theta)\right).
\end{equation}
The operator $O_{h_t}$ is determined by the truncated series. Let $\tilde{O}_{h_t}$ denote the operator ${U^{\dagger}}^{\otimes t} (\theta) O_{h_t} U^{ \otimes t }(\theta) $. We then prove that the requirement of $\mathcal{S}$-invariant data label leads to the modified symmetry constraint equation,
\begin{equation}
    P\left(\mathcal{T}\big(\tilde{O}_{h_t}(\theta^*) \big) - \tilde{O}_{h_t}(\theta^*)\right)=0,
\end{equation}
where $P$ projects an observable in a Hilbert space $\mathcal{H}$ into the symmetric subspace of $\mathcal{H}$. The proof and the concrete forms of $O_{h_t}$ and $P$ are shown in the Appendix~\ref{appendix:Classical post-processing}.

{\textit{Connection to other results.}---The effect of SGGD is similar to incorporating the symmetrically transformed data into the training dataset. A frequently used method to utilize symmetry is data enhancement~\cite{faris2002multiple, frieden2005image}, which appends the data generated by the symmetry transformation of the training set. In the example of 2D classification, the training data from the right-hand side (as shown in Fig.~\ref{fig:prl_twoDim}(a)) is in a sense transformed to the upper side since the SWAP symmetry transformation swaps the $x_0$ and $x_1$ coordinates of the input data.
In Fig.~\ref{fig:prl_twoDim}(d), the SGGD performs very well in the whole space except for the left-bottom part. Thus, the result of SGGD is similar to data enhancement in this case. }

{
However, data enhancement can lead to a significant increase in the size of the dataset, which introduces computational overhead. In contrast, SGGD does not increase the training set size but adds overhead to the evaluation of the cost function.  Since data enhancement creates more training data, the process of training models on the enhanced data may become more challenging~\cite{meyer2023exploiting, biwei2021dataaug}. 
An intuitive example is that, as the dimension increases, if we want to maintain a constant data density within a unit volume, the total data volume must grow exponentially with the dimension. The required quantity of data increases exponentially with the dimensions~\cite{bronstein2021geometric}. Linearly increasing the data size is insufficient in high-dimensional spaces. Consequently, resorting to the twirling technique becomes a more viable option in such scenarios because the SGGD achieves similar results without significantly increasing the number of training data points. Balancing the cost overhead of increasing the dataset size and the computational cost in the cost function can help to choose a suitable method.}

{The SGGD approach has a challenge in common with other penalty term approaches and we find a method to deal with this challenge. The challenge is that in principle there is a risk of the modification to the cost function impacting the training results negatively~\cite{bubeck2015convex, belegundu2019optimization}. This issue appeared in our numerical experiments of 2D classification. To alleviate this problem, we adjust the weight of the penalty term dynamically during training. We incrementally augment the weight of the penalty term to ensure the parameters are not over-guided by the symmetry constraint. This gradual increase in the size of the penalty term results in successful training that outperforms the non-guided training. (See details in the Appendix~\ref{appendix:Adjusting coefficient}.) }

\textit{Summary and outlook.--}
We impose symmetry constraints on the QNN's training, establishing a symmetry-guided gradient descent (SGGD) method. Our results show that SGGD can accelerate the training, improve the generalization ability, and remove vanishing gradients, especially when the training data is biased. Our method contributes understanding of how symmetry can be efficiently exploited in quantum machine learning and is widely applicable to tasks with either classical or quantum input data.

Apart from applying SGGD for enhancing QNN performance, further developments should be investigated. Optimal strategies should be explored for (i) varying the strength of the symmetry guidance during the training, and (ii) the norm used in defining the guidance~\cite{aggarwal2001surprising}. {It should also be investigated to what extent the method enhances the performance of more sophisticated forms of gradient descent such as the generalisation of gradient descent known as mirror descent~\cite{sbahi2022provably}.}The general method can also be adapted to guide circuit design more generally: one expects that good designs for a given problem should have a small symmetry penalty. 

\textit{Acknowledgements.--}We acknowledge inspiring discussions with Dong Yang and Jin-Long Huang. We acknowledge support from HiSilicon, the National Natural Science Foundation of China (Grants No.12050410246, No.1200509, No.12050410245), and the City University of Hong Kong (Project No. 9610623).

\bibliographystyle{ieeetr}
\bibliography{ref.bib}

\pagebreak
\onecolumngrid
\appendix

\section{ Hardware inefficiency of circuit ansatz symmetrization.}
\label{appendix:Hardware inefficiency of circuit ansatz symmetrization}
One approach commonly employed to leverage symmetry is modifying the circuits to adhere to the constraints imposed by symmetry, {which is referred to {\em ansatz symmetrization}}. An illustrative example is the gate symmetrization method described in the Refs.~\cite{meyer2023exploiting}.

We will first introduce the gate symmetrization procedure and then discuss the overhead that makes such symmetrization hardware-inefficient\cite{kandala2017hardware}. Assume that the ansatz comprises a set of gates denoted as $G$, given by
\begin{equation}
    G = \{ e^{i g_1 \theta_1}, e^{i g_2 \theta_2}, e^{i g_3 \theta_3}, \cdots  \}, 
\end{equation}
where the $g_i$ represents the generators of the ansatz. As an example, let's consider a hardware-efficient ansatz consisting of rotation gates $R_X(\theta)$ and CZ gates,
\begin{equation}
    \{ e^{i X \theta}, e^{-i A \frac{\pi}{4}}  \},
\end{equation}
where $A = I - Z_1 - Z_2 + Z_1Z_2$. The gate symmetrization technique twirls the set of generators to fulfill the symmetry requirement. For the previously mentioned hardware-efficient ansatz, the symmetrized circuit has the generators $\{ \T(X), \T(A) \}$. If the output of the quantum circuit is invariant under the $\{I, \mathrm{SWAP} \} $-operation, the twirled generators can be expressed as
\begin{align*}
    \T(X_1) =& \frac{1}{|\SSS|} \sum_{S\in \SSS} S X_1 S^\dagger\\
    =&\frac{1}{2}[X \otimes I+\mathrm{SWAP}(X \otimes I) \mathrm{SWAP}]\\
    =& \frac{1}{2}(X_1 + X_2)
\end{align*}
and $\T(A) = A$. In the main text, we define that $\SSS$ is a discrete symmetry group. By utilizing these twirled generators, a new set of gates is constructed,
\begin{equation}
    \{ e^{i (X_1+X_2) \frac{\theta}{2} }, e^{-i A \frac{\pi}{4}}  \}.
\end{equation}
This new gate set is then used to assemble a new quantum circuit that satisfies the required symmetry. 
Consequently, a quantum circuit that preserves symmetry is constructed through the twirling process. The procedure of constructing a symmetry-preserving QNN from the generators is referred to as ansatz symmetrization.

Now, we will analyze a specific case in which the symmetrization process will introduce an extra burden to the quantum device. Considering we apply a ZZ gate $e^{i\theta Z_1Z_2}$ on the quantum circuit, and the output of QNN is invariant under the permutation operators.  
The symmetrization of this gate gives the generator
\begin{equation}
    \T(Z_1Z_2) = \frac{1}{n!} \sum_{\sigma\in S} \sigma Z_1Z_2 \sigma^\dagger = \frac{2}{n(n-1)} \sum_{i\neq j} Z_iZ_j,
\end{equation}
where $\sigma$ is a permutation operator, and $S$ consists all possible permutations.
The realization of variational gates corresponds to the generator $\T(Z_1Z_2)$ are
\begin{equation}
    e^{i\T(Z_1Z_2)\theta}  = \prod_{i,j} e^{ i \frac{2\theta}{n(n-1)} Z_iZ_j},
    \label{eqn:sm_1}
\end{equation}
which are pairwise connected ZZ gates over all qubits. {Any two qubits in the circuit have a ZZ gate that connects them.}
{Thus, in this example, the twirled gate in Eq.~\eqref{eqn:sm_1} requires $\frac{n(n-1)}{2}$ gates to implement, which significantly increases the number of gates to implement. }

Furthermore, in scenarios where the quantum device is limited to performing only adjacent two-qubit gates, non-adjacent gates must be decomposed into a sequence of adjacent gates. {Most of the pairwise ZZ gates are not adjacent gates. If we want to apply the gate described in Eq.~\eqref{eqn:sm_1}, we need to decompose non-adjacent ZZ gates to adjacent basic gates. Now, we evaluate the cost of such decomposition. Firstly, we decompose the ZZ gate into basic gates because current quantum devices can not directly apply it, }  
\begin{equation}
\scalebox{1.0}{
\Qcircuit @C=1.0em @R=0.2em @!R { \\
	 	&\qw & \ctrl{1} & \qw                               & \ctrl{1} & \qw \\
	      &\qw & \targ    & \gate{\mathrm{R_Z}\,(\theta)} & \targ & \qw \\
\\ }}.
\end{equation}
Then, the ZZ gate on $i$-th and $j$-th qubits need to be decomposed to adjacent gates. We could use SWAP gate to decompose a $Z_i Z_j$ gate to adjacent ZZ gate,
\begin{equation}
\scalebox{1.0}{
    \Qcircuit @C=0.9em @R=0.1em @!R { \\
	 	\lstick{i} & \ctrl{3}& \qw & \ctrl{3} & \qw & \qw &  & & \qw & \qswap & \qw & \qswap & \qw & \qswap & \qw & \qswap & \qw & \qw  \\
	 	\lstick{i+1}& \qw    & \qw & \qw & \qw & \qw &  \push{\rule{.3em}{0em}=\rule{.3em}{0em}} & & \qw & \qswap \qwx[-1] & \qswap & \qswap \qwx[-1] & \qw & \qswap \qwx[-1] & \qswap & \qswap \qwx[-1] & \qw & \qw\\
	 	\lstick{i+2}& \qw    & \qw                           & \qw   & \qw & \qw &  & & \qw & \qw & \qswap \qwx[-1] & \ctrl{1} & \qw & \ctrl{1} & \qswap \qwx[-1] & \qw & \qw & \qw\\
	 	\lstick{i+3}& \targ  & \gate{\mathrm{R_Z}\,(\theta)} & \targ & \qw & \qw &  & & \qw & \qw & \qw & \targ & \gate{\mathrm{R_Z}\,(\mathrm{1})} & \targ & \qw & \qw & \qw & \qw\\
\\ }}
\end{equation}

To decompose a single $Z_i Z_j$ gate, excluding the case where $|i-j| = 1$, we require the addition of $2(2|i-j| -3)$ adjacent swap gates. The twirling of one ZZ gate needs $\frac{1}{3} \left(2 n^3-9 n^2+7 n\right)$ adjacent gates to achieve, 
\begin{align}
    (n-1) + 2(n-2) + 6(n-3) + 10(n-4) + \cdots + (4(n-1)-6)
    =\frac{1}{3} \left(2 n^3-9 n^2+7 n\right) 
    \sim  \mathcal{O}(n^3) ,
\end{align}
which costs too many gates for the current noisy quantum devices.

In addition to gate symmetrization, other approaches aimed at constructing symmetry-preserving blocks within QNN encounter a similar challenge. Typically, the symmetry operators employed are global unitaries that span multiple qubits. Thus, as we have shown, it becomes evident that only a global gate can maintain global symmetry. A global gate is usually hard to achieve in quantum devices~\cite {krantz2019quantum, huang2020superconducting}. 

\section{Symmetry constraint equation}
\label{appendix:Symmetry constraint equation}

In this section, we aim to give the proof of Eq.~(4) in the main text. Recall the symmetry requirement
\begin{equation}
    \tr(\rho(x) S^\dagger\widetilde{O}(\theta^*)S ) = \tr(\rho(x) \widetilde{O}(\theta^*)),
    \label{eqn: appendix 3}
\end{equation}
where $S$ are symmetry operations, $x$ is classical input data, and $O$ is the observable. We want to prove that the Eq.~\eqref{eqn: appendix 3} is equivalent to the symmetry constraint equation,
\begin{equation}
    \T(\widetilde{O}(\theta^*)) = \widetilde{O}(\theta^*),
    \label{eqn: appendix symmetry constraint equation}
\end{equation}
if the data set $\{\rho(x)\}$ span the space of states.

{In fact, Eq.~\eqref{eqn: appendix 3} are many equations because there are many symmetry operations. This equivalence merges many equations into one neat equation, which makes it easy to apply in the cost function.}

\begin{proof}
    Firstly, we give the case of discrete groups $ \SSS$. To show that Eq.~\eqref{eqn: appendix 3} and Eq.~\eqref{eqn: appendix symmetry constraint equation} are equivalent, we need to show that the solution sets of the two equations are the same. Precisely, we need to prove $A\subset B$ and $B \subset A$, where
    \begin{align}
        A:=&\{U \mid  \tr(\rho(x) S^\dagger\widetilde{O}(\theta^*)S ) = \tr(\rho(x) \widetilde{O}(\theta^*))\}\\
        B:=&\{U \mid  \T(\widetilde{O}(\theta^*)) = \widetilde{O}(\theta^*)\}.
    \end{align}
    Recall that $\Tilde{O}(\theta) := U^\dagger(\theta) O U(\theta) $. 
    
    We begin with the proof of $A\subset B$. $U\in A$ means
\begin{align*}
    \sum_{S\in \SSS} \tr(\rho(x)S^\dagger \Tilde{O}(\theta) S ) =& |\SSS|  \tr(\rho(x) \Tilde{O}(\theta))\\
    \tr(\rho(x)\frac{1}{|\SSS|}\sum_{S\in \SSS} S^\dagger \Tilde{O}(\theta) S ) =&   \tr(\rho(x) \Tilde{O}(\theta))\\
    \tr(\rho(x) \left(\frac{1}{|\SSS|}\sum_{S\in \SSS} S^\dagger \Tilde{O}(\theta) S - \Tilde{O}(\theta) \right)) =& 0 , ~~\forall \rho(x) \in \mathcal{H}\\
\end{align*}

Because $\{\rho(x)\}$ span the space, $\tr(\rho(x) A) = 0 ~~\forall \rho(x)$ implies $A = 0$. Therefore, the following equation holds,
\begin{equation}
    \frac{1}{|\SSS|}\sum_{S\in \SSS} S^\dagger \Tilde{O}(\theta) S - \Tilde{O}(\theta) = 0 .
    \label{eqn: SOS = O}
\end{equation}
Thus, Eq.~\eqref{eqn: appendix symmetry constraint equation} holds. It means $\forall U \in A$, $U\in B$, thereby $A\subset B$. 
Now, we prove that $B\subset A$. For any $U\in B$ and any $S_0\in \SSS$, we have 
\begin{align}
    \tr( \rho(x) S_0^\dagger\Tilde{O}(\theta) S_0 ) =& \tr( \rho(x) S_0^\dagger \left(\frac{1}{|\SSS|}\sum_{S\in \SSS} S^\dagger \Tilde{O}(\theta) S\right) S_0 ) \nonumber \\
    =&\tr( \rho(x)  \left(\frac{1}{|\SSS|}\sum_{S\in \SSS} S_0^\dagger S^\dagger \Tilde{O}(\theta) SS_0\right)  ) \nonumber \\
    =& \tr( \rho(x)  \left(\frac{1}{|\SSS|}\sum_{S\in \SSS}  S^\dagger \Tilde{O}(\theta) S\right)  ).
    \label{eqn: P1 prove1}
\end{align}
In the first line, we just substituted the Eq.~\eqref{eqn: appendix symmetry constraint equation} into the left-hand side. The last line is based on the rearrangement theorem of a group. 
Again, using Eq.~\eqref{eqn: appendix symmetry constraint equation}, the Eq.~\eqref{eqn: P1 prove1} becomes 
\begin{equation}
    \tr( \rho(x) S_0^\dagger\Tilde{O}(\theta) S_0 ) = \tr( \rho(x)  \left(\frac{1}{|\SSS|}\sum_{S\in \SSS}  S^\dagger \Tilde{O}(\theta) S\right)  ) = \tr( \rho(x) \OO).
\end{equation}
Thus, for all $U$ in $B$, $U$ is also in $A$, which means $B\subset A$. 
Combining the result $A\subset B$, we get that $A=B$.

For continue groups (Lie groups), the proof is the same because the rearrangement theorem works for both discrete groups and continuous groups. 

\end{proof}

\section{Classical post-processing}
\label{appendix:Classical post-processing}

In this section, we discuss how to derive the penalty term of SGGD to the cost function when there is a non-trivial post-processing on the output of the QNN.

Our analysis presented in this section greatly increases the applicability of SGGD, as in many cases post-processing cannot be avoided---the output of QNN is not the class label of the data but the probability of the data being in each class. For instance, assuming the input state $\ket{\psi} = \ket{010}$ has a label of $1$. A QNN that yields $f(\ket{010}) = 0.9$ signifies that the model assigns high probability to the state $\ket{010}$ correctly corresponding to label 1, rather than label the state $\ket{010}$ as $0.9$. Requiring the invariance of the output of QNN with respect to symmetry is too restrictive, as it requires $f(\ket{010}) = f(S\ket{010})=0.9$ for all the symmetry operator $S$. However, if $f(\ket{010}) = 0.9$ but $f(\ket{010})=0.8$, then the state $\ket{010}$ will again be assigned label 1, and the symmetry of the data label is respected.
Suppose the post-processing that takes the probability to the label is the step function $\Theta$, where 
\begin{equation}
    \Theta(x) = \begin{cases}
        1 \text{ ~~~if } x\geq 0\\
        -1 \text{ ~~if } x<0
    \end{cases}.
\end{equation}
Under this classical post-processing, the label invariance is expressed as
\begin{equation}
    \Theta(f(\rho)) = \Theta(f(S \rho S^\dagger)) 
\end{equation}
{ Here, {$f(\cdot)$} represents the output of the QNN, which provides the probability of the data belonging to each class.}


{Imposing symmetry to post-processing benefits the training and performance of the QNN.  As pointed out above, imposing the symmetry constraint on the output of QNN adds more constraints to the parameters than is required. If the circuit ansatz is not capable of completely satisfying the symmetry, requiring the QNN output to be invariant to the symmetry operation may reduce the accessible region of the parameters to a very small set that does not contain the optimal parameter $\theta^*$. This potentially makes a ansatz incapable of satisfying the requirements $f(\rho) = f(S \rho S^\dagger)$} 
. In this case, the optimal parameters $\theta^*$ are not in the symmetry-conserved space $\Omega'$. The symmetry penalty term likely produces a negative effect, { which we observed in our numerical experiments---the penalty term without post-processing pushes the convergence to a sub-optimal parameter with worse accuracy.} 

{ Here, we show that symmetry penalty terms can be designed with classical post-processing considered. Our method of adding symmetry penalty terms incorporating classical post-processing greatly increases the applicability of SGGD. Now, we introduce the detailed construction of the symmetry penalty term where there is classical post-processing}. Suppose the post-processing function $h$ is an analytical function that could be expanded by a Taylor series. Truncate the Taylor series in $t$-order, we get an approximate value of $h(x)$,
\begin{equation}
    h(x) = a_0 + a_1 x + a_2 x^2 + \cdots + a_tx^t + \mathcal{O}(x^{t+1}),
\end{equation}
where $a_j$ is the $j$-th order derivative of $h$, $a_j = \frac{d^jh}{dx^j}(0)$, and $N$ is the order of $\rho(x)$. Applying the post-processing to the model output $f_\theta(x)$, we get
\begin{equation}
    h (f(\rho(x))) = a_0  + a_1 \tr( U(\theta) \rho(x) U^\dagger(\theta) O) + a_2\tr( U(\theta) \rho(x) U^\dagger(\theta) O)^2 + \cdots.
    \label{eqn: classical post-processing object function 0}
\end{equation}
Note that 
\begin{equation}
    \tr( U(\theta) \rho(x) U^\dagger(\theta) O)^k = \tr((U(\theta) \rho(x) U^\dagger(\theta))^{\otimes t} ~ O ^{\otimes k} \bigotimes I^{\otimes  t-k}),
\end{equation}
the Eq.~\eqref{eqn: classical post-processing object function 0} becomes
\begin{equation}
    h (f(\rho(x))) = \tr((U(\theta) \rho(x) U^\dagger(\theta))^{\otimes t} \sum_{k = 0}^t a_k  O ^{\otimes k} \otimes I^{\bigotimes  t-k}).
    \label{eqn: classical post-processing objectfunction}
\end{equation}

The deduction of this equation is in the supplemental material. For simplicity, define the operator $F_{(h,t)}(O)$ as $F_{(h,t)}(O):= \sum_{k = 0}^t a_k  O ^{\otimes k} \bigotimes \mathbbm{1}^{\bigotimes  t-k}  $. 
While considering the classical post-processing, the symmetry invariant requirement reads as 
\begin{equation}
    \tr((U(\theta^*) S \rho S^\dagger U^\dagger(\theta^*))^{\otimes t} F_{(h,t)}(O)) = \tr((U(\theta^*) \rho U^\dagger(\theta^*))^{\otimes t} F_{(h,t)}(O))
    \label{eqn: classical post-processing symmetry}
\end{equation}
for $\forall S \in \SM$ and optimal parameter $\theta^*$.
Notice that Eq.~\eqref{eqn: classical post-processing symmetry} is not equivalent to the condition 
\begin{equation}
 \mathcal{T}(\tilde{O}_F(\theta^*)  ) = \tilde{O}_F(\theta^*),    
 \label{eqn: CPP no true}
\end{equation}
where $\tilde{O}_F(\theta^*) = U^\dagger(\theta^*)^{\otimes t} F_{(h,t)}(O) U(\theta^*)^{\otimes t} $. Here, we present an example to support the statement. 
Because the observable $A =  I \bigotimes X - X \bigotimes I$ make 
$ \tr(\rho(x) \otimes \rho(x) A) = 0 $
 for $\forall \rho(x) \in \mathcal{H}$, 
the twirled operator $\mathcal{T}(\tilde{O}_F(\theta^*)  ) = \tilde{O}_F(\theta^*) + A$ could also satisfy the symmetry invariant condition of Eqn. \eqref{eqn: classical post-processing symmetry}.

This counter-example implies a way of symmetry penalty term adjustment. 
Rather than the restriction in Eq.~\eqref{eqn: classical post-processing symmetry}, the constraint of optimal operator $\tilde{O}_F(\theta^*)$ turns to 
\begin{equation}
    \mathcal{T}(\tilde{O}_F(\theta^*)  ) - \tilde{O}_F(\theta^*) \in \mathcal{A}_t,
    \label{eqn: classical post-processing constraint: in A}
\end{equation}
where $\mathcal{A}_t := \{O  | \tr(\rho(x)^{\otimes t} O) = 0 , \forall \rho(x) \in \mathcal{H} \}$ is the solution space. 
Due to the Schur-Weyl duality~\cite{etingof2011introduction}, the solution space is equivalent to the symmetry subspace of the whole Hilbert space. Precisely, let the projection operator by
\begin{equation}
        P = \frac{1}{t!} \sum_{\sigma\in \Xi} \sigma,
    \end{equation}
    where $\Xi$ is the $t$-element permutation group. 
The solution space $\mathcal{A}_t$ is equivalent to the image of the projection $P$. Thus, Eq.~\eqref{eqn: classical post-processing constraint: in A} induces that 
 \begin{equation}
        (I-P)\Big(\mathcal{T}(\tilde{O}_F(\theta^*)  ) - \tilde{O}_F(\theta^*)\Big)=0.
        \label{eqn: proposition 2}
    \end{equation}
Eq.~\eqref{eqn: proposition 2} draws out the modified symmetry constraint equation. The modified symmetry penalty term is $g_{(h,t)}(\theta) = \norm{(I-P)\Big(\mathcal{T}(\tilde{O}_F(\theta)  ) - \tilde{O}_F(\theta)\Big)}$.

\begin{figure}
    \centering
    \includegraphics[width = 0.6\linewidth]{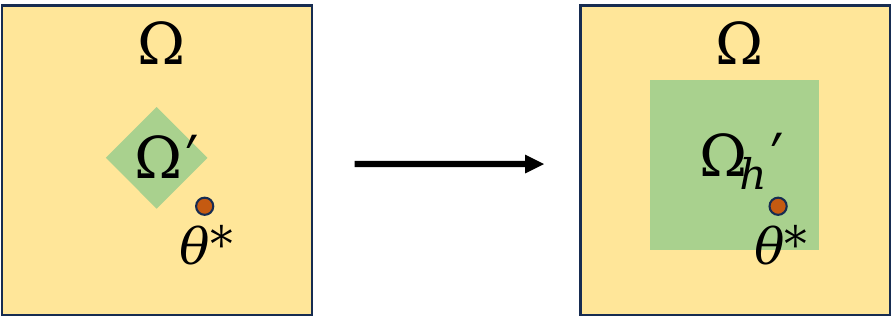}
    \caption{Illustration of the effect of classical post-processing. The green area of the left-hand side square means the solution space of the symmetry constraint equation. If non-trivial classical post-processing is present, the imposed constraint may become excessively stringent such that the optimal parameter could not be located in the green solution space. The modification in Eq.~\eqref{eqn: proposition 2} loses the constraint so that the space with symmetry, the green area on the right-hand side, is enlarged.}
    \label{fig: optimal space change under CPP}
\end{figure}

Fig.~\ref{fig: optimal space change under CPP} shows the effect of the adjustment. Because a solution of equation $g(\theta)=0$ must be a solution of equation $g_{(h,t)}(\theta)$ while a solution of equation $g_{(h,t)}(\theta)$ may not be a solution of equation $g(\theta)=0$, the symmetry preserved parameter space expand after the adjustment. If a QNN is applied to classical post-processing, adjustments must be made. As we mentioned before, a QNN model may not satisfy Eq.~\eqref{eqn: CPP no true} when the label is $\SSS$-invariant rather than the model output is $\SSS$-invariant. In this case, the optimal parameters lie out of $\Omega'$, which is the solution space of the symmetry constraint equation~\eqref{eqn: CPP no true}. The adjustment of the symmetry penalty term makes the symmetry-preserved space, $\Omega_h'$ large enough to contain the optimal parameter, where $\Omega_h'$ is the solution space of the modified symmetry constraint equation~\eqref{eqn: proposition 2}.

\section{Adjusting the coefficient of the symmetry penalty term to gain better training performance.}
\label{appendix:Adjusting coefficient}
 In this section, we present a strategy to adjust the coefficient of the symmetry penalty term to achieve better training of a QNN.
In the research of quantum supervised learning, one of the frequently chosen cost functions is the mean squared error between data labels and predictions, 
\begin{equation}
    c(\theta, \mathcal{X}) = \frac{1}{|\mathcal{X}|} \sum_{x \in \mathcal{X}} \Big( f_\theta(\rho(x)) - f(\rho(x))  \Big)^2,
\end{equation}
where $\mathcal{X}$ is the data set. 
We add the symmetry penalty term to this cost function,
\begin{equation}
    c_g(\theta, \mathcal{X}) = c(\theta, \mathcal{X}) + \lambda g(\theta).
    \label{eqn: cost function added guidance}
\end{equation}
Here, the $\lambda$ controls the intensity of the symmetry penalty term. 

With cost function $c_g$, the data is updated in the following way during gradient descending,
\begin{equation}
    \theta^{(i+1)} = \theta^{(i)} + \frac{d c(\theta, \mathcal{X})}{d\theta} \Big |_{\theta^{(i)}} \eta + \lambda\frac{d g(\theta)}{d\theta} \Big |_{\theta^{(i)}} \eta,
\end{equation}
where $\theta^{(i)}$ is the parameter of the $i$-th iteration, and $\eta$ is the learning rate. The extra term $\lambda\frac{d g(\theta)}{d\theta} \Big |_{\theta^{(i)}} \eta$ contributes a force that leads to the symmetry-preserved space $\Omega'$. What is more, the symmetry penalty term $g(\theta)$ would reach the minimum value $0$ when parameters $\theta$ are in the $\Omega'$. Hence, the $g(\theta)$ guides the parameters toward $\Omega'$, and it vanishes when $\theta\in\Omega'$.

\begin{figure}
    \centering
    \includegraphics[width = 0.8 \linewidth]{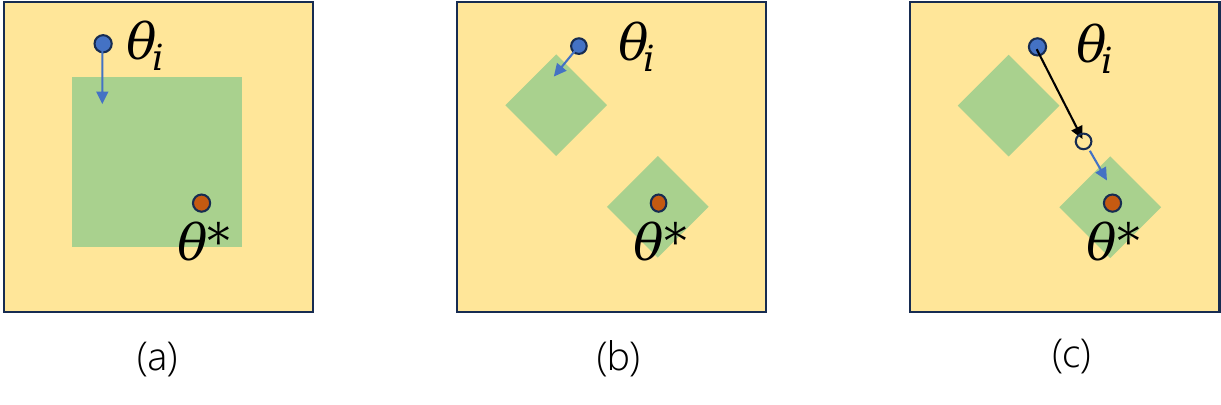}
    \caption{The misleading problem of the symmetry penalty term can be relieved by adjusting the weight of the penalty term dynamically. The yellow square represents the whole parameter space, and the green area represents the space with symmetry. (a) shows a working case that the symmetry can successfully guide the parameter $\theta_i$ closer to the optimal point $\theta^*$, where the subscript $i$ means the parameter of $i$-th iteration. (b) shows that the symmetry guides the parameter into the wrong place. In this case, there are two unconnected symmetry-preserving spaces, and the symmetry penalty term will attract the training to the wrong side if the initial configuration is close to the upper left green region. (c) shows the case when we apply the dynamic coefficient. The parameter will update without symmetry first, whose process is labeled by the black arrow. Then, the parameter will be updated with the help of symmetry. The dynamical coefficient could solve the misleading issue if we choose a proper strategy of changing the coefficient $\lambda$.}
    \label{fig: island, leading and misleading}
\end{figure}


The controlling coefficient $\lambda$ should be chosen carefully for practical applications. 
Notice that the symmetry penalty term only cares about symmetry rather than if the model makes the correct prediction. {If the symmetry penalty is assigned with a large weight at the beginning, then the symmetry may misguide the neural net to a wrong parameter configuration.} As shown in Fig.~\ref{fig: island, leading and misleading}, if the symmetry preserved space $\Omega'$ is not connected, there are several islands in $\Omega'$. When $\lambda$ makes $ \lambda\frac{d g(\theta)}{d\theta}  \gg \frac{d c(\theta, \mathcal{X})}{d\theta} $, the parameters tend to be trapped in one of the isolated regions where symmetry is satisfied. However, once the parameters are initialized close to a region that does not contain the optimal parameter, the training will be trapped in a local minimum, and performance will be bad. In this sense, we say that the symmetry misleads parameters. 

However, {{the symmetry misleading}} phenomenon could be relieved by a proper coefficient $\lambda$ picking strategy. Several efficient strategies are given here. The strategy uses dynamic intense controlling coefficient, which is shown in Fig. \eqref{fig: island, leading and misleading}. We could let the $\lambda$ be small at the beginning and let it become larger as the number of iterations increases. This strategy could be interpreted as one kind of pre-training. In the beginning, the $\lambda$ is negligible while $\frac{d c(\theta, \mathcal{X})}{d\theta} $ is dominant. At this time, the original cost function $c$ leads the parameters close to the optimal parameters $\theta^*$. After the parameters $\theta$ close to $\theta^*$, the $\lambda$ becomes larger and larger, and the symmetry penalty term begins to have an impact that cannot be ignored. Then, the misleading phenomenon disappears in this case because the model parameters are close enough to the optimal one. 

Another strategy is fixing an appropriate $\lambda$. If the lambda is bound by the following inequality in most places, the cost function will be the dominant term under most circumstances,
\begin{equation}
    \lambda \leq \abs{ \frac{dc}{dg}   }.
    \label{eqn: lambda bound}
\end{equation}
Because the bounded $\lambda$ makes $\lambda\frac{d g(\theta)}{d\theta}  \leq \frac{d c(\theta, \mathcal{X})}{d\theta} $ for most  $ \theta \in \Omega$. In this case, the symmetry penalty term just assists the original cost function $c$. Although its hard to select a subset $\Upsilon\subset \Omega$ that satisfies $\mu(\Upsilon) = (1-\epsilon)\mu(\Omega) $, where $\mu$ is a measure in $\Omega$ and $\epsilon$ is a small number, we could select the $\lambda$ empirically. In the numerical experiments we did, taking $\lambda = \frac{1}{3} \max(\abs{ \frac{dc}{dg}   })$ is small enough to avoid such misleading. 


Both strategies have their advantages. The utilization of an unsuitable dynamic coefficient can have detrimental effects on performance, compromising the effectiveness of the approach. On the other hand, developing a well-designed dynamic algorithm necessitates a considerable investment of time and effort. While the second strategy presents practical convenience in its implementation, it is important to note that the efficacy of a fixed parameter $\lambda$ employed in this approach falls short when compared to the performance achieved by a dynamically adjusted parameter $\lambda$ that exhibits favorable behavior.

\section{Supplemental Numerical Results}
\label{appendix:supplement numerical}

In the following, we first show another example that utilizes the first kind of symmetry-guided gradient descent $c_1$ to classify two-qubit bitstrings into two categories, which we call the ``cat-dog example." Then, we provide two two-parameter QNNs for the Werner state classification case in our main content, along with a detailed calculation process. We also present the experiment details for the 2D classification task.

\subsection{Cat-dog example}

\subsubsection{Task formulation}

Consider categorizing all pure quantum states under the computational basis into two categories, which means states corresponding bit-strings with only $1$s are labeled as cats while others as dogs. Then, we choose the QNN as $f(\ket{\psi}) = \tr(\ketbra{\psi}{\psi}\text{Z})$, where $\text{Z}$ is chosen as the observable as we only need to consider state amplitudes. Finally, we select $0$ as the threshold to identify these two classes, which means states with $f(\ket{\psi})\leq0$ would be classified as cats while others as dogs. 


This task could be considered from the view of classical logic circuits, where we process classical bit-strings. These tasks are transformed into implementing logical expressions using logical circuits, like $Y = \overset{-}{A}B + A\overset{-}{B}$ for the two-bit case, where Y is the logical circuit output, and A and B are bits inputs. This means we can always find suitable quantum circuits to solve these tasks.


As only the number of $0$s or $1$s would influence their category, the general form of the symmetry group here could be formulated as
\begin{equation}
	H_{\text{cat}} = \biggl\{\prod_{(i,j)}\mathrm{SWAP}_{i j}~|~ i \not=j ~,~ i,j  = 1,2,...,n \biggr\}.
\end{equation}
\subsubsection{Two-qubit case}
We first consider the simplest $n=2$ case. According to the two-bit case classical logical expression, which corresponds to the functional expression of an XOR gate, we could use circuits with only two qubits and detect the output value of the second qubit.


The classification task could be achieved by $X$ gates and CNOT gates in $2$ qubits since $f$ is a binary function. Thus, the quantum circuit that achieves the task must consist of $X$ gates and CNOT gates. We use  $\mathrm{X}^{\cos(\theta)^2}$ and $\mathrm{CNOT}^{\cos(\theta)^2}$ to switch the gates. If $\theta = 0$, the gates are open, else if $\theta = \frac{\pi}{2}$, the gates are closed. 
The QNN we used here is 
\begin{equation}
    U(\theta, \phi) = {\rm{(I}} \otimes {\rm{X)}}_{}^{{{\cos }^2}(\phi )}{\rm{CNOT}}_{{\rm{A}} \to {\rm{B}}}^{{{\cos }^2}(\theta )},
\end{equation}
where $\text{A}$, $\text{B}$ denotes the first and second qubit in our circuit, and $\to$ means $\text{A}$ controls $\text{B}$. After the input state evolves by the QNN, we measure the second qubit in $\text{Z}$-basis. 

\begin{figure}[h]
    \centering
    \includegraphics[width=0.4\textwidth]{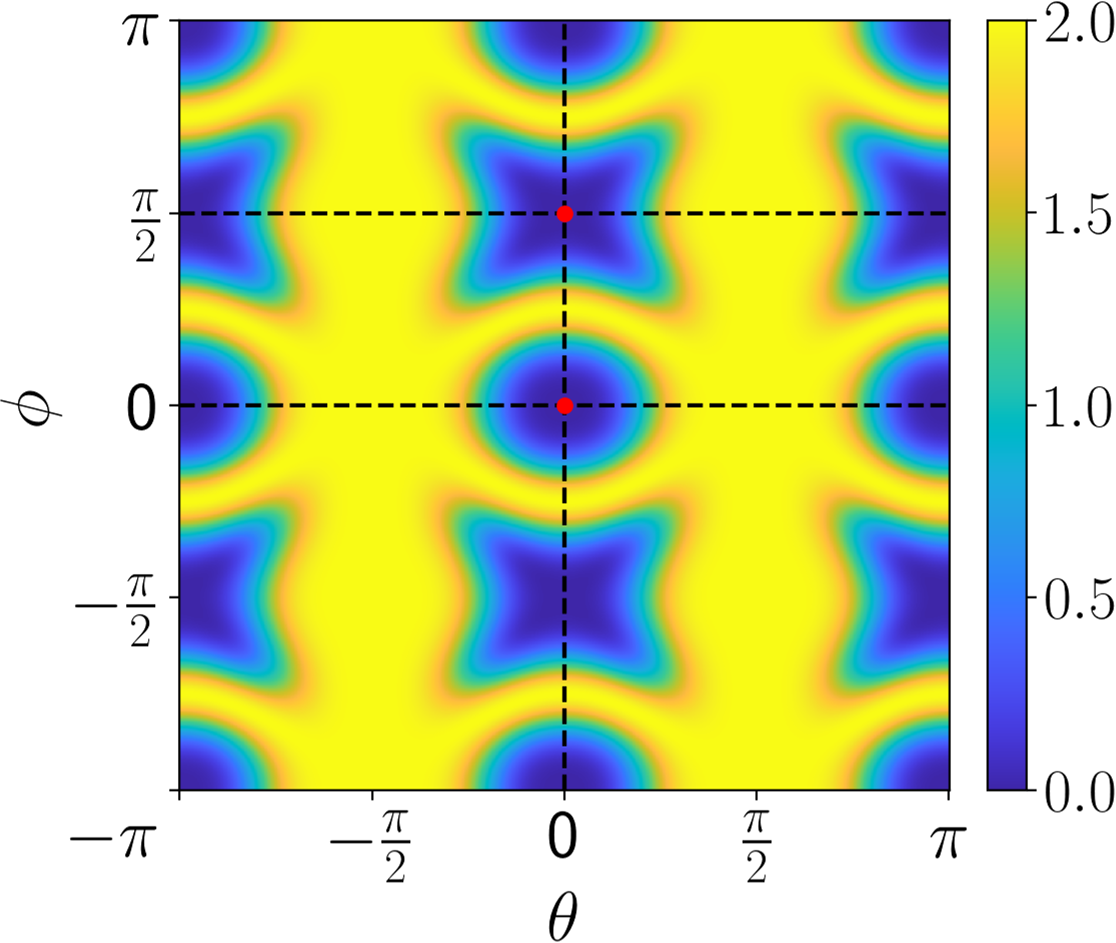}
    \caption{symmetry penalty term landscape of 2-qubit cat-dog example. The highlighted region signifies the area subjected to a substantial penalty. The red dots are the minimum of the penalty term. }
    \label{fig:cat_ix_sg}
\end{figure}

The symmetry group in this task is $H_1 = \{\text{SWAP}, \text{I}\}$ because the number of $1$s is invariant under permutation. With the symmetry group, we could calculate the symmetry penalty term
\begin{equation}
    g(\theta, \phi) = \norm{U^\dagger(\theta, \phi) Z_1 U(\theta, \phi) - \T(U^\dagger(\theta, \phi) Z_1 U(\theta, \phi)) }.
\end{equation}
The heat map of the symmetry penalty term $g(\theta, \phi)$ is shown in Fig.~\ref{fig:cat_ix_sg}. Notice that $\cos^2(\theta) = \cos^2(\theta+\pi)$. There are two in-equivalent minima $(\theta,\phi) = (0, \pi/2)$ and $ (0, 0)$ of the penalty term. If $\ket{01}$ and $\ket{10}$ are labeled $1$, the optimal parameter is $(0,0)$; else if $\ket{01}$ and $\ket{10}$ are labeled $0$, the optimal parameter is $(0, \pi/2)$. Thus, the two minima of the symmetry penalty term correspond to the two labeling strategies. 


\begin{figure}[t]
    \centering
    \includegraphics[width=0.9\textwidth]{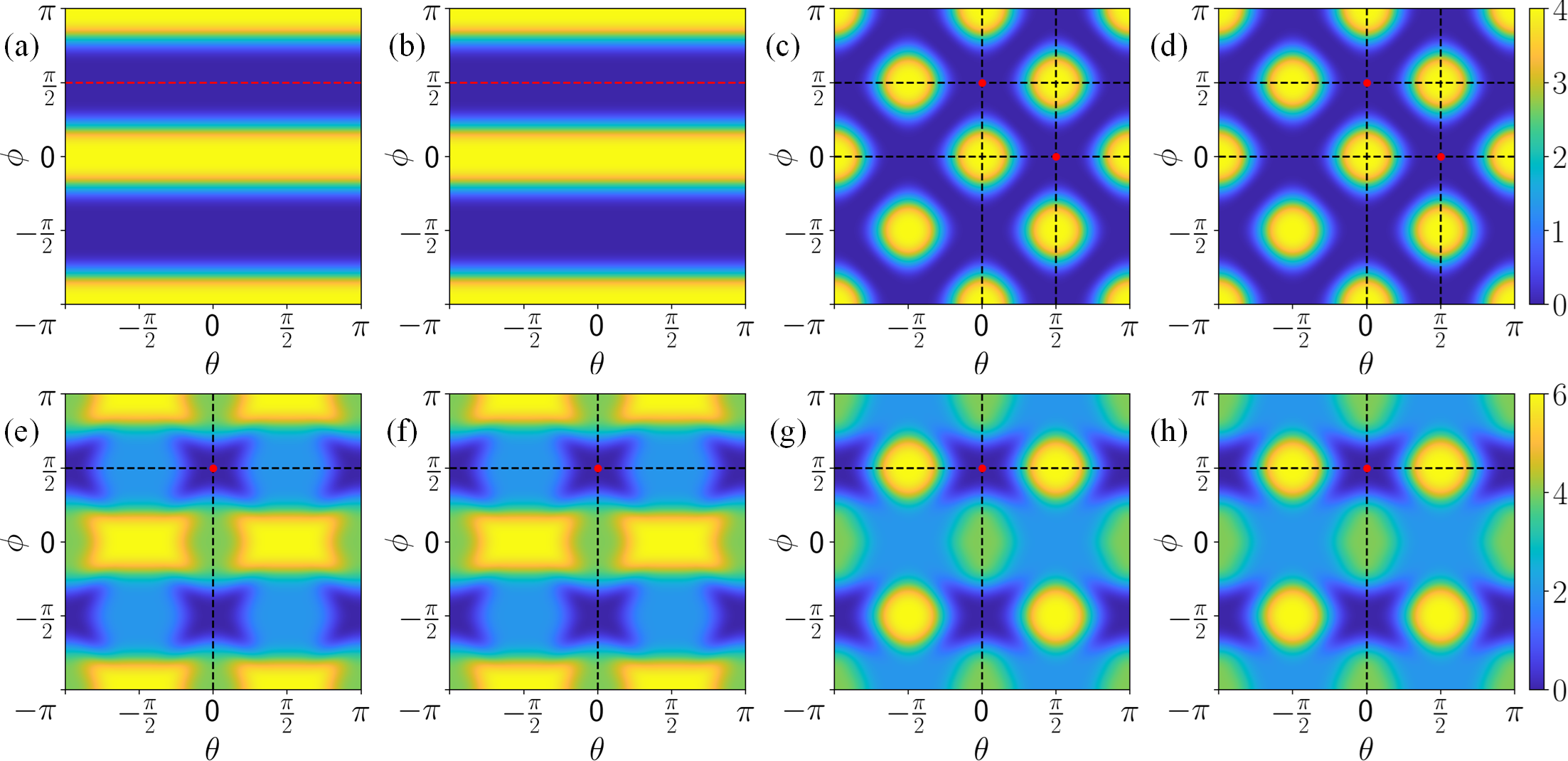}
    \caption{Loss landscapes before and after adding symmetry penalty term under all inputs}
    \label{fig:cat_ix_comp}
\end{figure}

Generally, there would be a discrepancy between theoretical and empirical errors, as losses or training would be biased under different training samples, which would then lead to over-fitting. This could be verified by the heat maps in our case, as shown in \eqref{fig:cat_ix_comp}, where we plot loss heat maps under different single inputs. In this case, under the default loss function, samples 00 and 01 would only train the $\phi$ parameter, while using either 10 or 11 would have two possible local minima, where one leads to the optimal point but another does not. But after adding a symmetry penalty term with a suitable guidance weight, we could see that all landscapes would have only one local minimum, and they all are exactly located at the optimal point.


We summarize the above results in \eqref{tab:catDog-2qubits}, where we can see that after introducing the symmetry penalty term, we could complete the task using only half samples (even only one) while needing four in the default setting. This comes from the guidance that strengthens the optimal points but alleviates parts that do not hold the symmetry this task requires.

\begin{table}[h]
\centering
\caption{Comparison among different cost function settings for 2-qubits cat-dog example}\label{tab:catDog-2qubits}
\renewcommand{\arraystretch}{1.3}
\begin{tabular}{|C{2cm}|C{3cm}|C{3cm}|C{4cm}|}
\hline
 & Symmetry penalty term & Loss function & Loss function + penalty term \\ 
\hline
Full Data & Half $\cross$ Half $\checkmark$ & $\checkmark$ & $\checkmark$ \\ 
\hline
Half Data & Half $\cross$ Half $\checkmark$ & Half $\cross$ Half $\checkmark$ & $\checkmark$ \\ 
\hline
\end{tabular}
\end{table}

This is an example showing that SG could help us deal with over-fitting. As for more qubits cases, it would be hard to show improvement easily, and the circuit shows some properties that could be further investigated.

\subsection{Entanglement classification }
\label{appendix:Entanglement classification}
\subsubsection{Task formulation and analysis}

Let us consider the example using QNN for Werner state entanglement witness. Werner states could be classified into entangled or separable states according to a given parameter $p$,
\begin{equation}
    \rho_{\mathrm{W}}=\frac{2-p}{6}\textbf{I}+\frac{2p-1}{6}\text{SWAP}.
\end{equation}
We formulate the task as finding a suitable observable that could detect whether a Werner state is separable or entangled according to its expectation value. In this task, we construct the objective function as $f(\rho)=\tr(U(\theta)\rho U(\theta)O)$. Then, we classify states according to their trace outputs, where we take states with $f(\rho)\leq0$ as separable states and states with $f(\rho)<0$ as entangled states.


Here, we work on the two-qubit case for illustration. A two-qubit Werner state could be constructed by $\rho_{\text{Werner}}=t\frac{I}{4}+(1-t)|\phi^-\rangle\langle\phi^-|$, where the factor $t$ ranges from 0 to 4/3. 2/3 is the split point of two kinds of states when $t\leq2/3$ provides separable states and $t>2/3$ provides entangled states.

This time, the symmetry group should be the whole set of single qubit unitary operators. We leverage tools provided in~\cite{mele2023introduction} (see Equation 48 in~\cite{mele2023introduction}) to calculate the twirled observable. This corresponds to calculating the second moment of a given observable $O$. In the two-qubit case, $\mathcal{T}(O)$ could be calculated as
\begin{equation}
    \mathcal{T}(O)=c_{\mathbb{I},O}I+c_{\mathbb{F},O}\mathbb{F},
\end{equation}
where $\mathbb{F}=\text{SWAP}$, $c_{\mathbb{I},O} = \frac{\tr(O) - 2^{-1} \tr(\mathbb{F}O)}{2}$, and $c_{\mathbb{F},O} = \frac{\tr(\mathbb{F}O) - 2^{-1} \tr(O)}{3}$.

\subsubsection{Two-qubit case 1}

First, we try the circuit form $U(\theta, \phi) = {\rm{CNOT}}_{{\rm{B}} \to {\rm{A}}}^{{{\cos }^2}(\phi )}{\rm{CNOT}}_{{\rm{A}} \to {\rm{B}}}^{{{\cos }^2}(\theta )}$. Then, we could get the landscape of symmetry penalty term where only one local minimum exists, as shown in \eqref{fig:werner_ix_comp} .

\begin{figure}[t]
    \centering
    \includegraphics[width=0.4\textwidth]{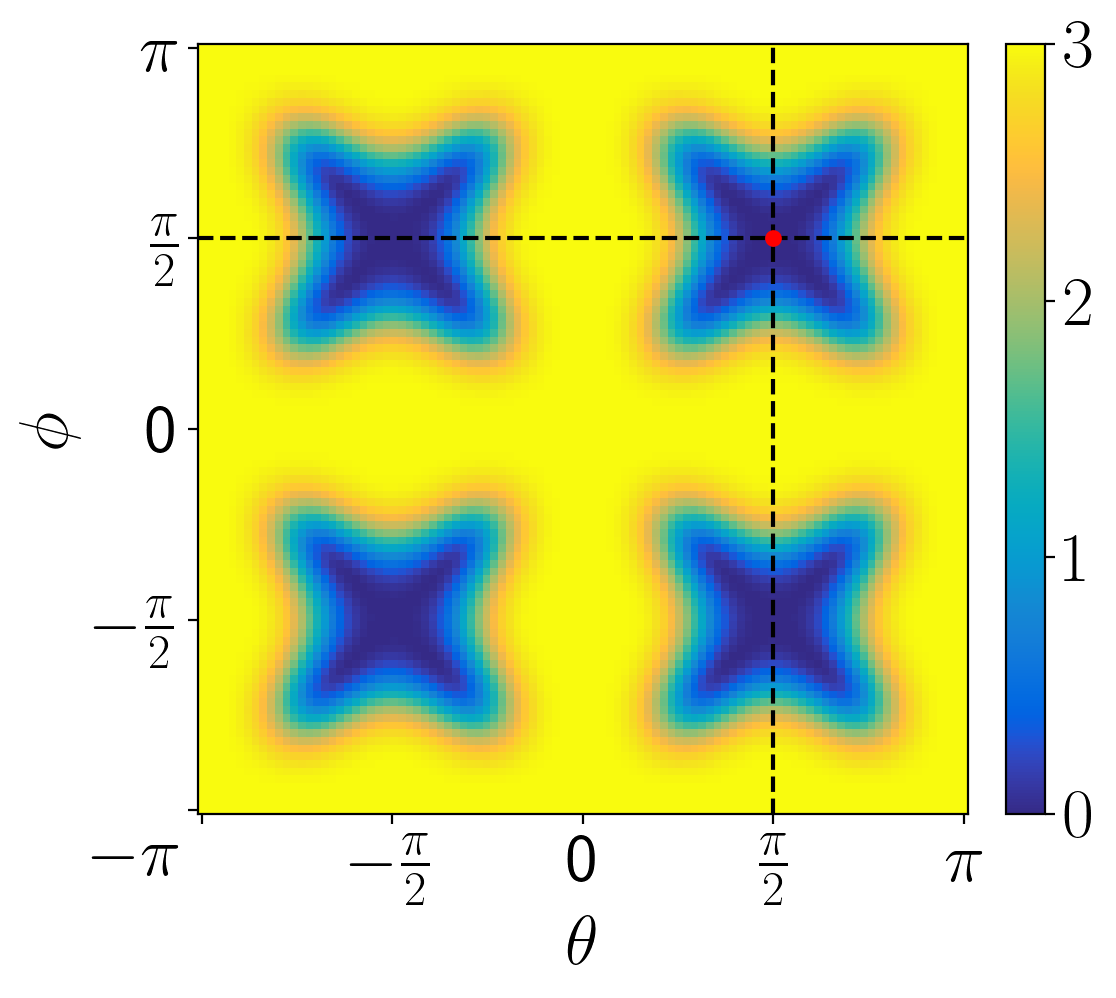}
    \caption{symmetry penalty term landscape of 2-qubit Werner example 1}
    \label{fig:werner_cnot2_sg}
\end{figure}

\begin{figure}[h]
    \centering
    \includegraphics[width=1\textwidth]{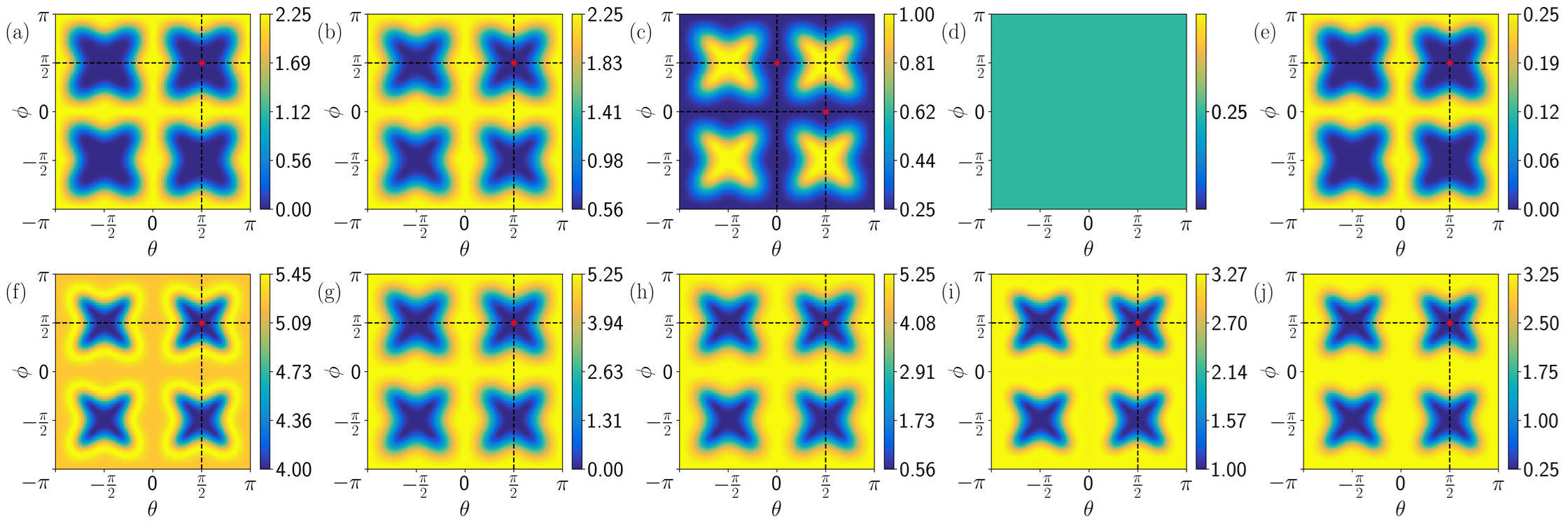}
    \caption{Loss landscapes before and after adding symmetry penalty term under selected input samples}
    \label{fig:werner_cnot2_comp}
\end{figure}

These samples could be separated into four groups based on the shape of their loss landscapes, though different samples having similar shapes would show exactly the same values. We select one sample from each class and show their loss landscapes in \eqref{fig:werner_cnot2_comp}. These four classes are entangled states ($0\leq t<2/3$ states), $t = 1$ state, separable states with $t<1$, and separable states with $t>1$. There is a split in separable states, as when gates in the given gates are not all-off, the trace output would be a constant value of $1/2$. 

From their loss landscapes, we could see that when using the default loss function, we need to select samples to train the model perfectly, as using samples with shapes in $(c)$ and $(d)$ would not lead to the overall optimal point. However, after adding the symmetry guidance term, we could train the model using any sample we like. This result is similar to the above two-qubit cat-dog case and shows that the symmetry penalty term could help biased samples less influence the optimization. But there is only one local minimum in this case, which is rare in normal tasks, so we consider a more complex one in the following.

\subsubsection{Two-qubit case 2}
This time, we use the same circuit in the above cat-dog example. We find that the model would output $1-0.5t$ when a single IX gate turns on, while all other circuits expect the identity circuit to output a constant value of $1/2$.

\begin{figure}[h]
    \centering
    \includegraphics[width=0.4\textwidth]{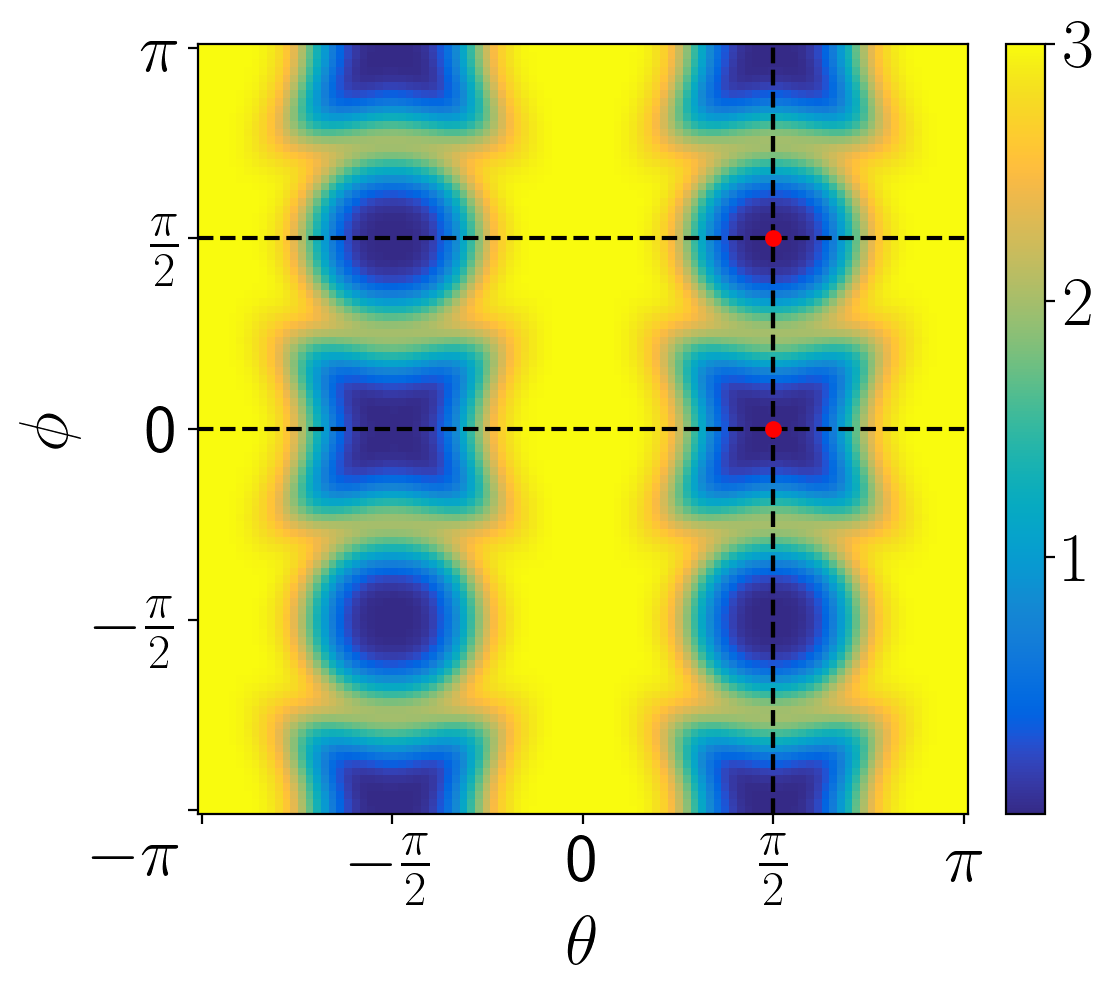}
    \caption{symmetry penalty term landscape of 2-qubit Werner example 2}
    \label{fig:werner_ix_sg}
\end{figure}

\begin{figure}[h]
    \centering
    \includegraphics[width=1\textwidth]{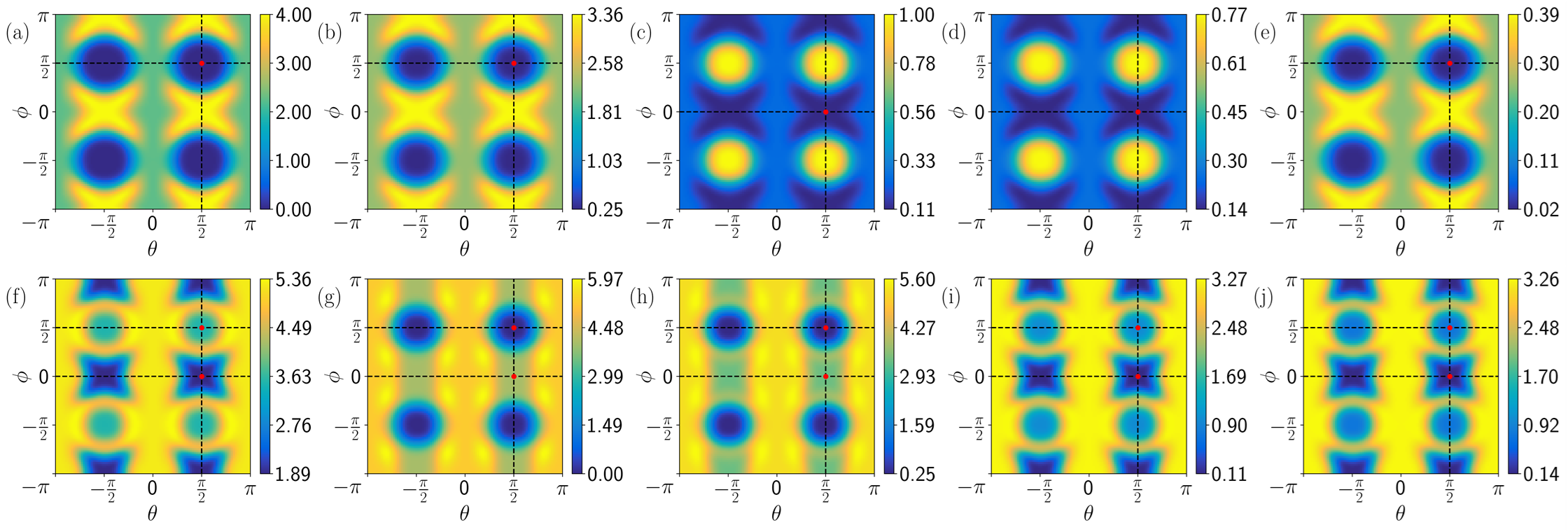}
    \caption{Loss landscapes before and after adding symmetry penalty term under selected input samples}
    \label{fig:werner_ix_comp}
\end{figure}

The symmetry penalty term now changed to have two optimal points, as shown in \eqref{fig:werner_ix_sg}, though we only want the $(\pi/2,\pi/2)$, or the all-gates-off one. Samples could also be categorized into four examples as above. 


In this case, as samples and their loss landscapes are shown in \eqref{fig:werner_cnot2_sg}, we can see that we still need to select samples for training perfectly in the default case. After adding the guidance term, though the improvement is not as significant as in the above two examples, new local minima were added for samples that would not train the model well in the default case. This means adding the guidance terms would help us alleviate sample-biased over-fitting in a probabilistic way. 

\subsection{Circuit and experiment details for 2-D classification}
\label{appendix:circuit ansatz}
\begin{figure}[H]
    \centering
    \includegraphics{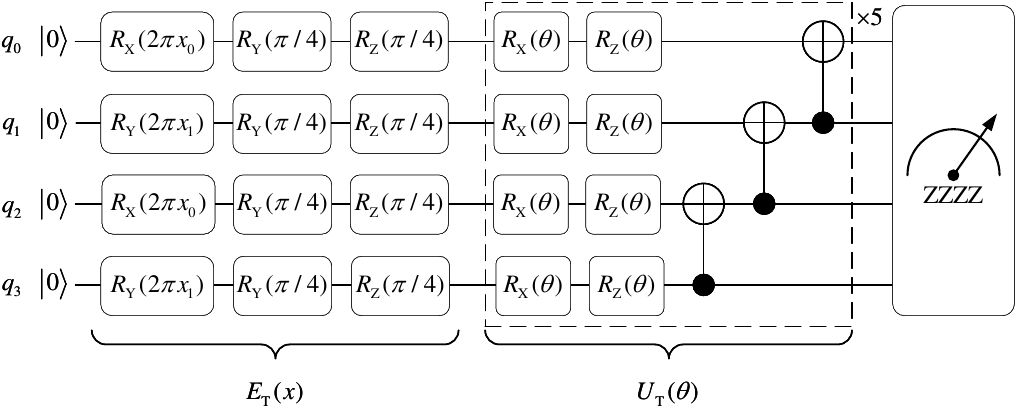}
    \caption{QNN for 2-D classical data classificaioin}
    \label{fig:sup_2d_qnn}
\end{figure}

The circuit we used for 2D classical data classification is shown in Fig.~\ref{fig:sup_2d_qnn}.

The coding is implemented via the PennyLane library~\cite{bergholm2018pennylane}. The gradient descent uses the Adam optimizer~\cite{kingma2014adam} with a learning rate of $0.01$ for both training with $c_0$ and $c_2$. Training data are randomly generated samples from each class, with 140 samples drawn from each area, as shown in the main text, Fig.3(a).  After the training, we evaluate the performances on samples uniformly selected from the whole data space. The final overall accuracy increased from $69.93\pm 4.51\%$ to $89.62\pm4.61\%$ after adopting our $c_2$ function in 21 runs of experiments. The classification results are shown in the main text, Fig.3(c) and (d).  Codes and data for our numerical experiments can be found on Github (https://github.com/Fragecity/SymmetryQNN).
\end{document}